\documentclass[runningheads]{llncs}
\usepackage[T1]{fontenc}
\usepackage{graphicx}
\usepackage{booktabs}
\usepackage[misc]{ifsym}

\usepackage{graphicx}

\usepackage{amsmath}
\usepackage{amsfonts}
\usepackage{bbold}
\usepackage[ruled,vlined]{algorithm2e}

\usepackage{multirow}
\usepackage{bbm}
\usepackage{caption}
\usepackage{subcaption}

\usepackage{svg}
\usepackage{array, booktabs, boldline}
\usepackage{adjustbox}
\usepackage{longtable} 
\usepackage{colortab} 
\usepackage{colortbl}
\usepackage{arydshln}

\usepackage{array}
\usepackage{amsmath}
\usepackage{xspace}

\newcommand{\methodours}{\texttt{SBO}\xspace}

\newcommand{\userset}{$U=\{u_1,\ldots, u_{|U|}\}$\xspace}
\newcommand{\itemset}{$V=\{v_1,\ldots, v_{|V|}\}$\xspace}
\newcommand{\usergroup}{$U_g$\xspace}
\newcommand{\itemster}{$I_\text{Ster}$\xspace}
\newcommand{\userster}{$U_\text{Ster}$\xspace}
\newcommand{\itemgi}{$IGI$\xspace}
\newcommand{\userstersthres}{$\gamma$\xspace}
\newcommand{\userobfitems}{$C_u$\xspace}
\newcommand{\userstermean}{$U_\text{Ster}^\text{mean}$\xspace}
\newcommand{\userstermed}{$U_\text{Ster}^\text{median}$\xspace}
\newcommand{\obfratio}{$\rho$\xspace}
\newcommand{\userobfcand}{$X_u^\rho$\xspace}
\newcommand{\topstereosampling}{$TopStereo$\xspace}
\newcommand{\randsampling}{$Random$\xspace}
\newcommand{\sbosampling}{$SBsampling$\xspace}
\newcommand{\perblur}{\texttt{Perblur}\xspace}

\newcommand{\modelmultvae}{\texttt{\textsc{MultVAE}}\xspace}
\newcommand{\modellightgcn}{\texttt{\textsc{LightGCN}}\xspace}
\newcommand{\modelbpr}{\texttt{\textsc{BPR-MF}}\xspace}

\newcommand{\stratimputate}{\textit{imputation}\xspace}
\newcommand{\stratremoval}{\textit{removal}\xspace}
\newcommand{\stratweight}{\textit{weighted}\xspace}

\newcommand{\dataMlm}{\texttt{\textsc{Ml-1m}}\xspace}
\newcommand{\dataLFM}{\texttt{\textsc{LFM-2b-100k}}\xspace}
\newcommand{\dataorigLFM}{\texttt{\textsc{LFM-2b}}\xspace}

\newcommand{\eg}{e.\,g., }
\newcommand{\ie}{i.\,e., }

\newcommand{\bacc}{\text{BAcc}\xspace}

\newcommand{\ndcg}{\text{NDCG}\xspace}

\definecolor{lblue}{HTML}{BBEAF0}
\definecolor{dblue}{HTML}{BBEAF0}

\newif\ifworkinprogress
\workinprogresstrue

\ifworkinprogress
    \newcommand{\ms}[1]{\textcolor{olive}{\textbf{[Markus] #1}}}
    \newcommand{\dk}[1]{\textcolor{red}{\textbf{[Dominik] #1}}}
    \newcommand{\get}[1]{\textcolor{blue}{\textbf{[Gustavo] #1}}}
      \newcommand{\pmu}[1]{\textcolor{cyan}{\textbf{[Peter] #1}}}
    \newcommand{\mm}[1]{\textcolor{orange}{\textbf{[Marta] #1}}}
\else
    \newcommand{\ms}[1]{}
    \newcommand{\dk}[1]{}
    \newcommand{\get}[1]{}
    \newcommand{\pmu}[1]{}
     \newcommand{\mm}[1]{}
\fi

\begin{document}

\title{Making Alice Appear Like Bob: A Probabilistic 
Preference Obfuscation Method For Implicit Feedback Recommendation Models}
\toctitle{Making Alice Appear Like Bob: A Probabilistic Preference Obfuscation Method For Implicit Feedback Recommendation Models} 
\titlerunning{Making Alice Appear Like Bob}
%
\author{
Gustavo Escobedo\inst{1}(\Letter)\orcidID{0000-0002-4360-6921} \and
Marta Moscati\inst{1}\orcidID{0000-0002-5541-4919} \and
Peter Muellner\inst{4}\orcidID{0000-0001-6581-1945} \and
Simone Kopeinik\inst{4}\orcidID{0000-0002-6440-7286} \and
Dominik Kowald\inst{4}\orcidID{0000-0003-3230-6234} \and
Elisabeth Lex\inst{3}\orcidID{0000-0001-5293-2967} \and
Markus Schedl\inst{1,2}\orcidID{0000-0003-1706-3406} 
}
\tocauthor{
Gustavo Escobedo, 
Marta Moscati,
Peter Muellner,
Simone Kopeinik,
Dominik Kowald, 
Elisabeth Lex, 
Markus Schedl 
}
\authorrunning{G. Escobedo et al.}
%
\institute{
Johannes Kepler University Linz, Linz, Austria\\
\and
Linz Institute of Technology, Linz, Austria\\
\email{\{gustavo.escobedo,marta.moscati,markus.schedl\}@jku.at}\\
\and
Graz University of Technology, Graz, Austria\\
\email{elisabeth.lex@tugraz.at}\\
\and 
Know-Center GmbH, Graz, Austria\\
\email{\{pmuellner,skopeinik,dkowald\}@know-center.at}
}
\maketitle              
\begin{abstract}
Users' interaction or preference data used in recommender systems carry the risk of unintentionally revealing users' private attributes (e.g., gender or race). This risk becomes particularly concerning when the training data contains user preferences that can be used to infer these attributes, especially if they align with common stereotypes. 
This major privacy issue allows malicious attackers or other third parties to infer users' protected attributes. 
Previous efforts to address this issue have added or removed parts of users' preferences prior to or during model training to improve privacy, which often leads to decreases in recommendation accuracy. 
In this work, we introduce \methodours, a novel probabilistic obfuscation method for user preference data designed to improve the accuracy--privacy trade-off for such recommendation scenarios. 
We apply \methodours to three state-of-the-art recommendation models (i.e., BPR, MultVAE, and LightGCN) and two popular datasets (i.e., MovieLens-1M and LFM-2B). Our experiments reveal that \methodours outperforms comparable approaches with respect to the accuracy--privacy trade-off. Specifically, we can reduce the leakage of users' protected attributes while maintaining on-par recommendation accuracy.
\keywords{Recommender Systems \and Privacy \and Obfuscation \and Debiasing \and Implicit Feedback}
\end{abstract}

\section{Introduction}
Recommender systems (RSs) provide relevant content to their users, commonly based on large collections of users' historical interaction data with items, using collaborative filtering techniques. The historical data used for training of and inference in recommendation models consists of interactions of users with several items and hence represent the preference of each user. 
While such user-item interaction data is necessary to create an accurate recommendation model, it may also reflect inherent biases in user behavior, which are subsequently encoded or even amplified during model training.
For instance, users of music recommender systems from different countries and of different genders tend to prefer different artists and genres~\cite{DBLP:journals/ijmir/Schedl17,DBLP:journals/mta/KrismayerSKR19,lex2020modeling}, leading to a correlation between users' sensitive attributes and behavioral patterns encoded in their interaction data. 

As a consequence, this leads to two important risks: 
possible {privacy} breaches and stereotypical or even unfair recommendations. 
As for \textit{privacy} issues, users' protected information can be leaked when untrusted third parties get access to the users' interaction data~\cite{DBLP:journals/mta/KrismayerSKR19} or internal user representation of the model~\cite{DBLP:conf/recsys/VassoyLK23,ganhoer2022advmultvae}.
For instance, for a group of users that is highly correlated with a list of stereotypical items, private attributes~(\eg gender, occupation, or country) can be unveiled through malicious attacks on the model or the data~\cite{anelli2022adversarial_handbook}.
Concerning \textit{unfairness}, recommendation models trained on interaction data that is correlated with sensitive user attributes have been shown to impact the quality of recommendations across different user groups distinguished by these attributes~\cite{melchiorre2021gender_fairness}.

Both problems (privacy concerns and fairness issues) are intertwined because they originate from the correlations between users' interaction behaviors and their sensitive attributes.
To mitigate them, several privacy-enhancing methods have been introduced, targeting different stages of the recommendation model's training process~(pre-, in-, and post-processing)~\cite{Mullner2023DP-CF}.   Among the pre-processing methods, user preference obfuscation approaches have been proposed to impede malicious attacks that aim at the leakage of private user attributes before training. 
These approaches primarily consist of adding or removing carefully selected items from users' preference data and have specifically been applied to user-item matrices containing ratings~\cite{DBLP:journals/ipm/SlokomHL21,DBLP:conf/recsys/StrucksSL19}.

In the work at hand, we introduce \textit{Stereotypicality-Based Obfuscation} (\methodours), a {probabilistic user preference obfuscation} method to counteract inference attacks against private user attributes. 
Unlike existing methods, \methodours selects users and items to obfuscate in a probabilistic fashion, using novel stereotypicality metrics. 
This limits the number of users whose items require obfuscation and adjusts the selection probability of non-stereotypical items in the sampling process.
We demonstrate \methodours's performance in terms of recommendation utility and accuracy of an attacker that aims to unveil the users' gender. Experiments with three common recommendation algorithms---\modelbpr, \modellightgcn, and \modelmultvae---on two standard recommendation datasets from the movie and music domains---\dataMlm (MovieLens)~\cite{DBLP:journals/tiis/HarperK16} and \dataLFM (Last.fm)~\cite{Schedl2022LFM2b,melchiorre2021gender_fairness}---
showed a favorable accuracy--privacy trade-off of our method.

In the remainder of the paper, we review relevant previous work (Section~\ref{sec:related}), detail the proposed \methodours method (Section~\ref{sec:method}), present the setup of our evaluation experiments (Section~\ref{sec:experiments}), and discuss results (Section~\ref{sec:results}). Ultimately, we summarize our findings and provide an outlook (Section~\ref{sec:conclusions}).

\section{Related Work}\label{sec:related}
Related work belongs to two strands of research: privacy-aware RSs (Section~\ref{sec:related:pars}) and fairness in RSs through adversarial training (Section~\ref{sec:related:debias}). Both can be addressed by altering the user's input data to the RS or the model's latent user representations.
 
\subsection{Privacy-aware Recommender Systems}\label{sec:related:pars}

RSs typically expose their users to several privacy risks. 
For example, the disclosure of information that is used to train the recommendation model (e.g., interaction data)~\cite{xin2023user,hashemi2022data} to third parties, or the inference of information that is not used during model training but correlated with the training data (e.g., gender or age)~\cite{weinsberg2012blurme,zhang2022comprehensive}.

Various technologies have been employed to address users' privacy concerns, such as homomorphic encryption~\cite{kim2016efficient}, federated learning~\cite{lin2020meta,muellner2021robustness}, and differential privacy~\cite{reuseknn,mullner2024impact}. 
Homomorphic encryption aims to generate privacy-aware recommendations by utilizing encrypted user data~\cite{zhang2021privacy}. 
Federated learning operates under the principle that sensitive user data should remain on the user's device~\cite{anelli2021federank}.  
Lastly, differential privacy (DP) is used to counter privacy risks by incorporating a carefully tuned level of random perturbation into the recommender system~\cite{dwork2008differential}. 
Many works apply DP to protect a user's sensitive attribute. 
However, 
malicious parties can still scrutinize the generated recommendations to infer protected attributes~\cite{ekstrand2018privacy}. This is the case if non-sensitive interaction data correlates with the user's sensitive attributes and forms distinct patterns that can be uncoded.

For this reason, Weinsberg et al.~\cite{weinsberg2012blurme} suggest an approach that detects rating data that is indicative of gender and adds ratings for items indicative of the opposite gender to obfuscate a user's real gender. 
However, the authors regard the set of items in a user profile as the source of the privacy risk (i.e., the correlation with gender), and their approach leads to a severe drop in recommendation accuracy.
In contrast, in the work at hand, we regard the \textit{conjunction} of items in the user profile as the source of the privacy risk, i.e., the correlation of the user's behavioral pattern with gender stereotypes.
Additionally, we address the accuracy drop by applying our perturbation mechanism only to users whose behavioral patterns coincide with gender stereotypes.






\subsection{Fairness Through Adversarial Training in Recommendation}\label{sec:related:debias}
    In the context of RSs, protecting users' privacy often relates to concepts of user fairness~\cite{anelli2022adversarial_handbook,deldjoo2021adversarial_fairness,wu2023fairness,dwork2012fairness,zemel2013learning}---a topic of lively interest in research and public communities ~\cite{ekstrand2022handbook_fairness,Wang23SurveyFairness,deldjoo2022fairness}. A particular overlap of the two strands exists with so-called fairness through unawareness or fairness through blindness approaches, where "unfair" bias is mitigated by hiding the users' sensitive attributes in the model training process~\cite{verma2018fairness}. Thus, privacy and fairness can potentially be ensured if the users' data on protected/sensitive attributes is not encoded in the model. 
    
    In RS research, several works use adversarial learning as an in-processing technique~\cite{jin2023fairness} to generate feature-independent user embeddings. 
    For instance, to achieve counterfactual fairness, Li et al.~\cite{Li2021counterFactMultfairness} apply an adversarial learning module to enforce user embeddings to be independent of the protected attributes. Ganhör et al.~\cite{ganhoer2022advmultvae} and Vass\o{}y et al.~\cite{DBLP:conf/recsys/VassoyLK23} add adversarial training to autoencoder-based RSs (e.g.,~\cite{lacic2020using}) to remove the implicit information of protected attributes from latent representations of users. Wu et al.~\cite{wu2021adversarial_fairness} use adversarial learning to develop a RS based on two representations of the user: a representation that carries the biased information through sensitive attributes and a bias-free representation that only encodes user interests. Wu et al.~\cite{wu2021representation_fairness} develop a graph-based adversarial learning module to increase the fairness of recommendations. More similar to our work, Weinsberg et al.~\cite{weinsberg2012blurme} and Strucks et al.~\cite{DBLP:conf/recsys/StrucksSL19} use obfuscation to achieve privacy; Slokom et al.~\cite{DBLP:journals/ipm/SlokomHL21} show that obfuscation also impacts the fairness of recommendations, while Lin et al.~\cite{DBLP:conf/trustcom/LinLZWHL22} use obfuscation to debias gender from RSs. 
    In contrast to prior works, the work at hand introduces the usage of the user's attribute-specific stereotypicality of items for the probabilistic selection of the data to obfuscate.
    


\section{Methodology}\label{sec:method}
The core idea of the proposed \methodours method is to reduce the \textit{stereotypicality} of the users' preferences by applying item obfuscation (imputation and/or removal) at the user level. For this purpose, we first define an item stereotypicality score~(\itemster) based on the item's group inclination ($IGI$). The $IGI$ value indicates how likely it is that a user of a given group consumes a certain item. Then, we use the \itemster values to establish the user's stereotypicality~(\userster) from the interaction data, which enables us to determine 
each user's degree of stereotypicality concerning the group to which the user belongs. For instance, a male user who predominantly listens to male-associated music tracks will obtain a high user stereotypicality score. 
\userster is then used to identify suitable candidates for obfuscation according to a given threshold. For each candidate user selected for obfuscation, we sample a number of items proportional to a fixed percentage of the number of items the user interacted with and apply obfuscation operations on the sample.  

We formally present our method in the subsequent sections, focusing on obfuscating gender\footnote{In this work, we consider only two possible values of gender. However, we acknowledge that the assumption of binary gender is an over-simplification.} information because it is a common target for attacks. Note that our method can be easily adapted for other protected attributes. 
We start by defining the different stereotypicality scores for users and items. Then, we formulate \methodours with the supported sub-sampling and obfuscation strategies. 

\subsection{Item's Group Inclination}\label{sec:igi}
We split the set of unique users \userset in $k$ groups $\{U_g\}_{g=1}^{k}$, where $U_g \subset U $ and $\bigcap_{g=1}^k U_g =\emptyset $, based on the $k \ge 2$ mutually exclusive values of the categorical protected attribute $p$ associated with each user. In this work, we split the original set of users by their associated gender. 
Therefore, we define two groups, $U_m$ and $U_f$, corresponding to the male and female users, respectively.

Items present different degrees of association to different user groups. Therefore, for each element in the set of unique items \itemset, we define the item inclination towards the user group 
\usergroup as the fraction between the number of users in \usergroup that interacted with item $v$, and the total number of users in \usergroup. Therefore, given the set of observed interactions $L_\text{obs} \subset U \times V$, $IGI(v,U_g)$ is given by:  

\begin{equation}\label{eq:gen_item_incl}
    IGI(v,U_g) = \frac {|\{u : (u,v)\ \in L_\text{obs}\}|}{|U_g|} 
\end{equation}

\subsection{Item Stereotypicality}\label{sec:item_ster}
In order to determine if an item is a good candidate for obfuscation, we introduce the item stereotypicality~(\itemster), which relates the \itemgi values of the same item~$v$~(Eq.~\ref{eq:gen_item_incl}) across two user groups. The closer the values of inclination across groups, the closer to zero the value of \itemster. This also indicates that the items closer to the extremes are the most stereotypical ones. The definition of \itemster and its dependence on $U_g$ and $U_{g'}$ is given by:
\begin{equation}\label{eq:flip factor}
    I_\text{Ster}(v, U_g, U_{g'})= \frac{ IGI(v,U_g) - IGI(v,U_{g'})}{\max \{ IGI(v,U_{g}), IGI(v,U_{g'}) \}}
\end{equation}
Therefore, \itemster$(v,U_g,U_{g'})$\ =\ --  \itemster$(v,U_{g'},U_g,)$. 
 
\begin{figure}[b]
    \centering
        \begin{subfigure}{0.5\textwidth}
        \centering
        \includegraphics[width=0.7\linewidth]{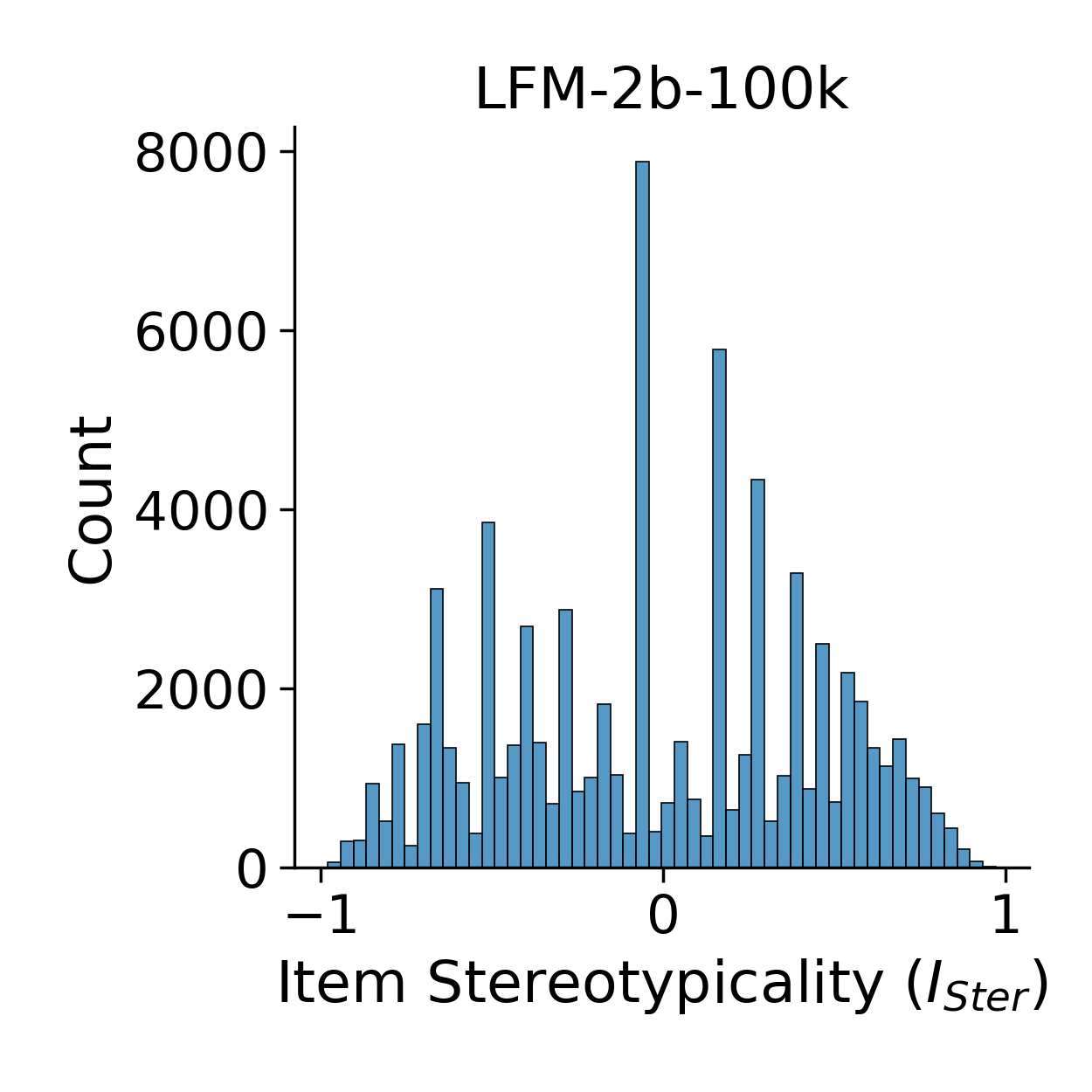} 
        \end{subfigure}\hfill
        \centering
        \begin{subfigure}{0.5\textwidth}
        \includegraphics[width=0.7\linewidth]{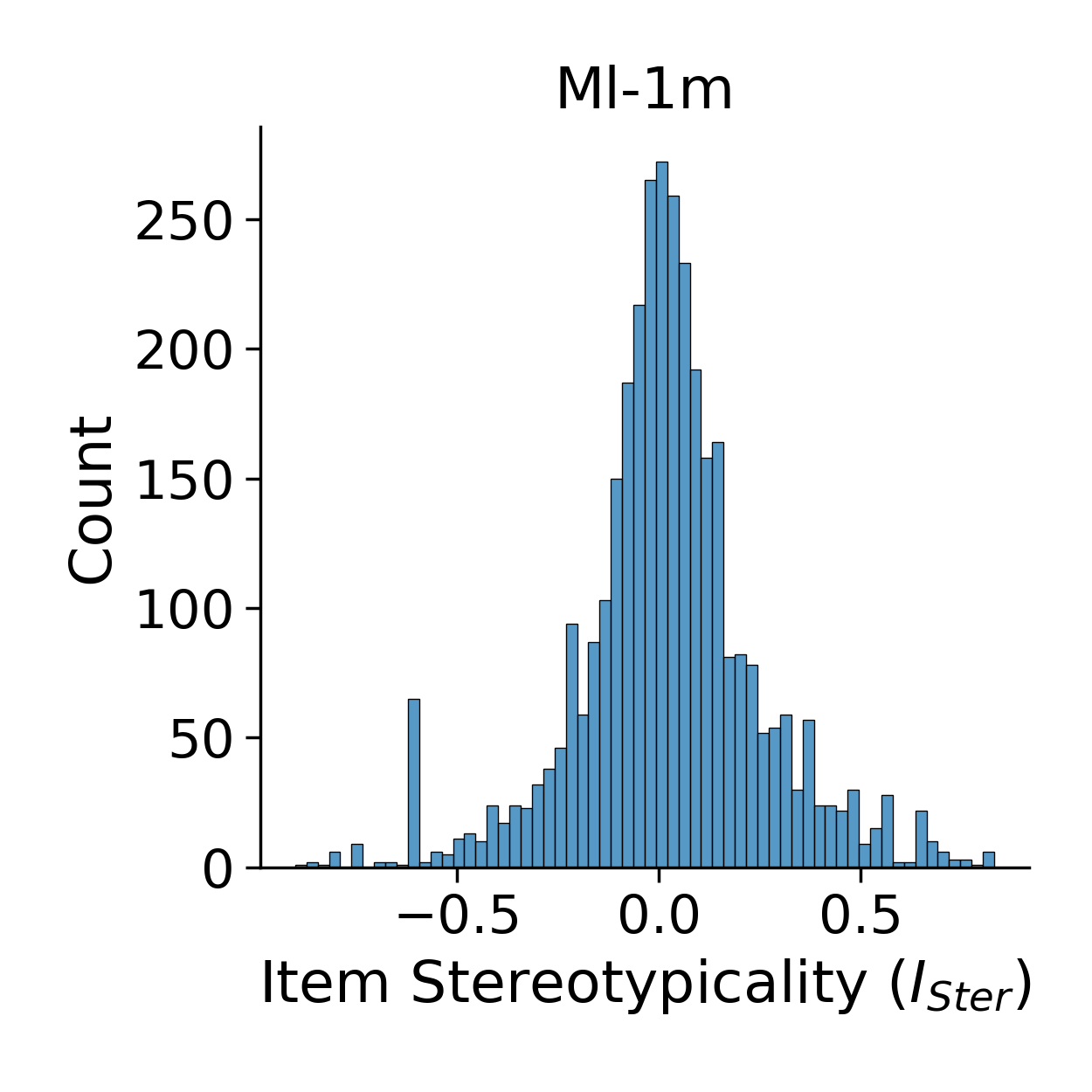} 
        \end{subfigure}
   
    \caption{Distribution of item stereotypicality~\itemster$(v,U_g,U_{g'})$ with $U_g=U_m$ and $U_{g'}=U_f$ over the  items of the \dataLFM (left) and \dataMlm (right) datasets.}
    \label{fig:lfm_item_ster}
\end{figure}

Figure \ref{fig:lfm_item_ster} shows the distribution of \itemster values over items for the \dataLFM and \dataMlm datasets, and for the users in the $U_m$ group, \ie setting $U_g=U_m$ and $U_{g'}=U_f$ in  Eq.~\ref{eq:gen_item_incl}. Whenever considering a user, we gather the corresponding \itemster values that match the value of the user-protected attribute. In addition, these values are calculated only for items that were consumed by at least one user in each user group.

\subsection{User Group Stereotypicality}\label{subsec:user_ster} 
Next, we introduce a measure of the target user's strength of preference towards group-biased or stereotypical items as defined in Subsection~\ref{sec:igi}.   
Given a user $u$ and the items in their profile $v \in X_u$, the user's preference towards stereotypical items is measured as the mean \userstermean or median \userstermed of the distribution of values of $I_\text{Ster}^u$ over the items in $X_u$. Throughout this paper, for simplicity, we refer to these scores as \userster for both definitions (mean and median), but separately explore the effects of both in our results. 

\begin{figure}[t]
    \centering
    \includegraphics[width=0.5\textwidth]{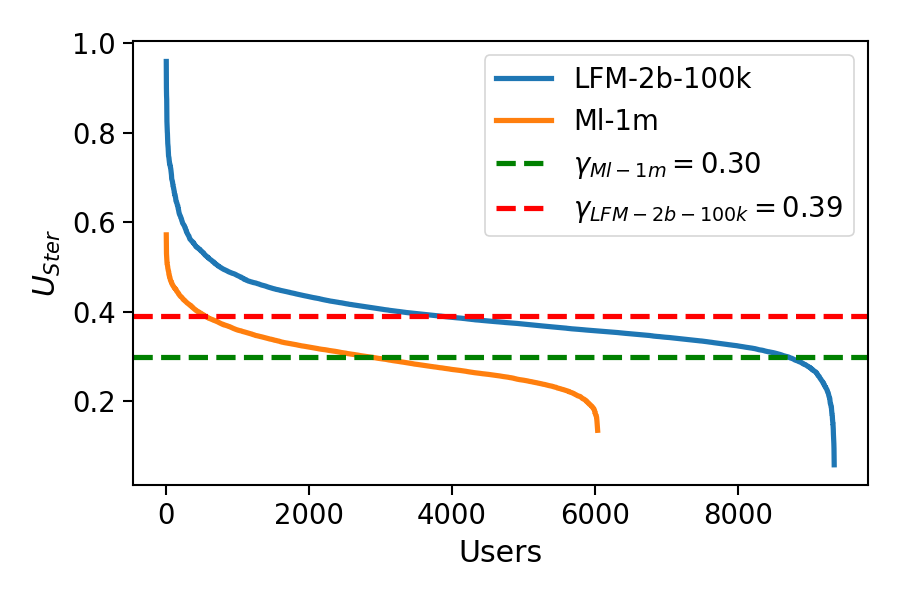}
    \caption{User group stereotypicality of users from the \dataLFM and \dataMlm datasets, with users in order of descending stereotypicality. The red dotted and green dotted lines indicate the selection threshold \userstermean used for \dataLFM and \dataMlm, respectively.}
    \label{fig:user_ster_thresh_dist}
\end{figure}

The \userster values are used to determine whether a user is to be considered \textit{highly stereotypical}. Therefore, we define the threshold of user stereotypicality \userstersthres as the mean value of all users' \userster scores. Users with \userster$\ge$\userstersthres are considered \textit{highly stereotypical} and hence selected as targets for obfuscation. Fig.~\ref{fig:user_ster_thresh_dist} shows the values of \userster of users from \dataLFM and \dataMlm in order of descending stereotypicality, as well as the thresholds \userstersthres.   



\subsection{Stereotypicality-based Obfuscation}\label{subsec:SBO} 
Our method \methodours consists of three main steps: 1) filtering users according to their \userster score; 2) sub-sampling candidate items; and 3) obfuscating the users' profiles. Below, we describe each step of the method separately and summarize them in  Alg.~\ref{alg:cap}.  
First, we compute the list $M_u$ of values of \itemster according to the user's gender label $g$ (each entry of $M_u$ representing a different item). 
Then, we compute the \userster values for each user and filter the users with scores higher than the threshold \userstersthres, which is considered as a hyper-parameter. Given an obfuscation ratio \obfratio, the item sub-sampling consists of selecting a set of obfuscation candidates \userobfcand for the user, containing at most \obfratio$\cdot \left|X_u\right|$ items. 
For this purpose, we define different sampling pools for the three different obfuscation strategies: \stratimputate, \stratremoval, and \stratweight. Specifically, the sampling pool for \stratimputate is $V-X_u$, and the sampling pool for \stratremoval is $X_u$; additionally, a weighted combination of these two is the sampling pool for \stratweight. The weight $\omega\le1$ decides on the number of items to select for imputation $\omega$ and for removal $1-\omega$ and is treated as a hyper-parameter.\footnote{We report results for $\omega=0.5$ only.}



\subsubsection{Sterotypicality-based Sampling.} 
To sample the items to select for obfuscation, \methodours first selects the items with the highest \itemster scores from the set of obfuscation candidates \userobfcand. Then, \methodours decides on the items to obfuscate by performing a Bernoulli trial on each of the selected items, with a success rate equal to the item's absolute \itemster values. Therefore, items that have high \itemster values are more likely to be obfuscated. 
The candidate items in \userobfcand in which Bernoulli trials were successful are inserted in the obfuscation items list \userobfitems, and then obfuscated. The Bernoulli trials are performed on each item independently to use the same \itemster values across all user profiles. 

Our aim is to obfuscate the items that are highly stereotypical in the user profile, therefore, when imputing unseen items, we choose items with the most negative \itemster scores (most counter-stereotypical for $u$'s gender). On the contrary, when removing items, we select the items with the most positive \itemster scores (most stereotypical for $u$'s gender). Following the same reasoning, we also define an additional baseline sampling strategy for comparison, \topstereosampling, where the items with the highest \itemster scores in the user profile are selected for removal and the most negative for imputation. In addition, we include the \randsampling strategy, which selects items uniformly at random from the user profile for removal and from the set of unexplored items for imputation. After having sub-sampled the list of candidate items  \userobfcand, we perform the selected obfuscation method using the \texttt{Obfuscate} on the user profile $X_u$ and the obfuscation strategy $m$. 


\begin{algorithm}[h!]
\DontPrintSemicolon
\caption{Stereotypicality-based Obfuscation}\label{alg:cap}
\SetKwData{EmptyList}{Empty List}
\SetKwData{True}{True}
\SetKwData{Male}{male}
\SetKwFunction{BernoulliTrial}
{BernoulliTrial}\SetKwFunction{SubSample}{SubSample}
\SetKwFunction{Obfuscate}{Obfuscate}
\SetKwInOut{Input}{input}\SetKwInOut{Output}{output}
\SetAlgoLined
\Input{List of items the user $u$ interacted with $X_u$,\\ 
User $u$'s gender label $g$, \\
List of unique items $V$,\\
User groups defined by gender $\{U_m$, $U_f\}$,\\ 
User stereotypicality threshold \userstersthres, \\
Obfuscation sampling ratio \obfratio,\\
Obfuscation strategy $m$
}
\Output{Obfuscated list of user $u$'s interactions $\Tilde{X_u}$}

\BlankLine
\tcp{Assigning user's stereotypicality}
$S_u \leftarrow$ $U_\text{Ster}(X_u)$\;
\tcp{User's obfuscation candidate items}
\userobfitems $\leftarrow$ $\{ \}$ \; 

$\tilde{X_u} \leftarrow$ $\{ \}$\;
\tcp{Defining the list of item stereotypicality values according to the user's gender label}
\eIf{$g ==$ \Male}{
    $M_u \leftarrow$ $\{I_\text{Ster}(v,U_m,U_f): v \in V\}$ \;
}
{
    $M_u \leftarrow$ $\{I_\text{Ster}(v,U_f,U_m): v \in V\}$ \;
}
\tcp{Evaluating the user for high stereotypicality}
\eIf{$S_u$ $\ge$ \userstersthres}{
    \tcp{Sub-sampling of candidate items to obfuscate}
    $X_u^\rho \leftarrow$ \SubSample{$V$, $X_u$, $\rho$,$\ m$}\;
    \For{$v \in X_u^\rho$}{
    $p\leftarrow \left|M_u(v)\right|$\;
    $c \leftarrow$\BernoulliTrial{$p$}\;
        \If{c==\True}{
        $C_u \leftarrow   C_u \cup \{v\}$\;
        }
    }
    \tcp{Performing obfuscation of the user profile $X_u$}
    $\tilde{X}_u \leftarrow$ \Obfuscate{$X_u,C_u,m$}\;
    }
{
 $\tilde{X}_u \leftarrow X_u$
}
\end{algorithm}


\subsection{Attacker Network}
As common in literature~\cite{Li2023IndistinguishableUsers,ganhoer2022advmultvae}, we use a simple feed-forward network as an attacker network. The network is trained on vector representations of the users' interaction data in a supervised manner to predict the private attributes from these representations. The successful prediction of the attribute implies that the current interaction data can reveal the values of the attributes. In our case, this network takes the user preference vectors as input and aims to predict the user's gender.  

\section{Experimental Setup}\label{sec:experiments}



\subsection{Datasets}

We run evaluation experiments on two popular datasets: \dataMlm~\cite{DBLP:journals/tiis/HarperK16}\footnote{\url{https://grouplens.org/datasets/movielens/1m/}} and 
\dataLFM,\footnote{A subset of \dataorigLFM~\cite{Schedl2022LFM2b,melchiorre2021gender_fairness}, derived by first selecting users with valid gender information, then randomly select $\sim100k$ unique items that adhere to 5-core filtering.  
} covering the movie and music domain, respectively. 

For the training of recommendation models, we apply 5-core filtering to each dataset, randomly select 20\% of each user's interactions, and leave them out as \textit{test} set. 
We apply the same split procedure on the remaining 80\% of interactions to generate the \textit{training} and \textit{validation} sets. For the attackers' training
, we perform 5-fold cross-validation over the set of unique users, leaving 20\% of them as test users in each fold, and report the average value of the evaluation metrics computed over the folds.

In order to perform obfuscation, we use the concatenation of the \textit{train} and \textit{validation} slices of the original datasets, then we slice the resultant set of interactions following the previously introduced procedure for both recommendation models and attacker networks. 


\begin{table}[t]
 \centering
\caption{Statistical description of datasets}\label{tab:datasets}
\begin{tabular}{lrrrr}
   
\toprule
Dataset &  Users (Male/Female) & Items & Interactions & Density \\\midrule
\dataMlm & $\quad$ 6,040 (4,331/1,709) & $\quad$  3,416 & $\quad\quad\quad$ 999,611 & $\quad$ 0.0484\\
\dataLFM & 9,364 (7,580/1,784)& 99,965 & 1,820,903  & 0.0019\\\bottomrule 
\end{tabular}

\end{table}

\subsection{Dataset Obfuscation}\label{subsec:dataset_obf}
The generation of obfuscated datasets is done before training the models with the following hyper-parameters: the user stereotypicality threshold \userstersthres is defined as the mean or median as described in Section~\ref{subsec:user_ster}, the obfuscation ratio parameter is set to \obfratio$=0.1$.\footnote{We also used \obfratio$=0.05$, obtaining similar 
results, for which we refer the reader to our supplementary material (Appendix \ref{sec:appendix1}).} We perform experiments for all the obfuscation strategies and sampling methods defined in Section~\ref{subsec:SBO}. We evaluate \methodours against a state-of-the-art obfuscation approach, \perblur, proposed by Slokom et al.~\cite{DBLP:journals/ipm/SlokomHL21}. Where available, we used the code provided by the authors\footnote{\url{https://github.com/SlokomManel/PerBlur}} and implemented the missing pieces of code. Specifically, we set \perblur's number of user neighbors to $50$ for \dataLFM and to $100$ for \dataMlm. From these neighbors, we collect the $200$ and $500$ most frequent recommended items for  \dataLFM and \dataMlm, and used them as personalized lists. Then, we follow the procedure described by Slokom et al.~\cite{DBLP:conf/recsys/StrucksSL19} for selecting the 50 most indicative items for each gender. We include in our results both the performance of \perblur with the imputation and with the removal method
.    

\subsection{Algorithms}
\paragraph{Recommendation Models.} Since the proposed method \methodours is largely independent of the recommendation algorithm as long as those are trained on implicit feedback, 
we carry out our experiments on a selection of well-established recommendation algorithms from different categories: matrix factorization (\modelbpr~\cite{DBLP:conf/uai/RendleFGS09}), neural network-based (\modelmultvae~\cite{LiangMultVAE2018}), and graph-based (\modellightgcn\cite{DBLP:conf/sigir/LighGCN}), hence demonstrating its performance across different types of RSs.
We train the RSs for 100 epochs with a learning rate of $0.001$ using the Adam optimizer with $512$ as batch size. We apply early-stopping with a patience of 10 epochs, using \ndcg as validation metric, computed for the top $10$ predicted (\ie recommended) items. The  embedding size of all models is set to 64
. We use the implementation of the RS models provided by the \texttt{RecBole}\footnote{\url{https://github.com/RUCAIBox/RecBole}} framework. Each model is evaluated with each of the dataset obfuscation parameters defined in Section~\ref{subsec:dataset_obf}.
\paragraph{Attacker Networks.} 
For the attacker networks, we define the architecture $A=[\left|V\right|,l,2]$ setting the number of nodes of the intermediate layer to $l=128$ for \dataMlm and to $l=256$ for \dataLFM. Each of the attackers is trained for 50 epochs using the Adam optimizer with 64 as batch size and $0.001$ as learning rate with a Cross-Entropy~(CE) minimization objective. In order to mitigate the imbalanced distribution of gender, we set proportional weights to each gender category in the CE objective. These networks are applied to all the configurations of parameters defined in Subsection~\ref{subsec:dataset_obf}.

\paragraph{Evaluation.}
To assess the recommendation performance, we report the Normalized Discounted Cumulative Gain~(\ndcg) 
for the top 10 recommended items. Additionally, we report the Balanced Accuracy~(\bacc) to assess the performance of the attacker networks. To ensure the reproducibility of our research, the implementation and complete configuration of our experiments can be found in our publicly available repository.\footnote{\url{https://github.com/hcai-mms/SBO}}

\section{Results and Discussion}\label{sec:results}
In this section, we describe our results, focusing first on the effect on the accuracy--privacy trade-off. We then delve into the effect of \methodours's different parameter configurations. 
\begin{table}[h!] 
\caption{Experimental results on the two datasets \dataMlm and \dataLFM. The scores in \textbf{bold} indicate the best scores across all models.}
\centering
\setlength{\tabcolsep}{6pt}
\begin{tabular}{lllrrrr}
\toprule
 &  &  &  & \modelbpr & \modellightgcn & \modelmultvae \\
 \cmidrule{5-7}
Dataset& Obf. Strat.  & Sampling &\multicolumn{1}{c}{\bacc$\downarrow$} & \multicolumn{3}{c}{\ndcg$\uparrow$} \\
 
\midrule
 & orignal & original & 0.5501 & 0.1135 & \textbf{0.1773} & 0.1483 \\
 \cdashline{3-7}
\multirow[c]{12}{*}{\dataLFM} & \multirow[c]{4}{*}{imputate} & PerBlur & 0.5522 & 0.1042 & 0.1561 & 0.1402 \\
 &  & Random & 0.5427 & 0.0990 & 0.1543 & \textbf{0.1607} \\
 &  & SBSampling & 0.5528 & 0.1209 & 0.1764 & 0.1513 \\
 &  & TopStereo & 0.5528 & 0.1209 & 0.1764 & 0.1513 \\
\cdashline{3-7}
 & \multirow[c]{4}{*}{remove} & PerBlur & 0.5471 & 0.1155 & 0.1764 & 0.1507 \\
 &  & Random & 0.5414 & 0.1070 & 0.1564 & 0.1324 \\
 &  & SBSampling & \textbf{0.5136} & 0.1138 & 0.1731 & 0.1441 \\
 &  & TopStereo & 0.5445 & \textbf{0.1224} & 0.1759 & 0.1518 \\
 \cdashline{3-7}
 & \multirow[c]{3}{*}{weighted} & Random & 0.5417 & 0.1055 & 0.1584 & 0.1504 \\
 &  & SBSampling & 0.5528 & 0.1209 & 0.1764 & 0.1513 \\
 &  & TopStereo & 0.5528 & 0.1209 & 0.1764 & 0.1513 \\
 \midrule
  & original & original & 0.6182 & 0.3445 & 0.3655 & 0.3650 \\
 \cdashline{3-7}
\multirow[c]{12}{*}{\dataMlm} & \multirow[c]{4}{*}{imputate} & PerBlur & 0.6156 & 0.3344 & 0.3581 & 0.3580 \\
 &  & Random & 0.5973 & 0.3389 & 0.3592 & \textbf{0.3718} \\
 &  & SBSampling & 0.8329 & 0.2866 & 0.3174 & 0.3154 \\
 &  & TopStereo & 0.8751 & 0.3111 & 0.3468 & 0.3499 \\
 \cdashline{3-7}
 & \multirow[c]{4}{*}{remove} & PerBlur & 0.6597 & 0.3437 & 0.3656 & 0.3657 \\
 &  & Random & 0.6076 & 0.2904 & 0.3116 & 0.3161 \\
 &  & SBSampling & \textbf{0.5664} & 0.3400 & 0.3608 & 0.3586 \\
 &  & TopStereo & 0.6124 & 0.3396 & \textbf{0.3679 }& 0.3650 \\
  \cdashline{3-7}
 & \multirow[c]{3}{*}{weighted} & Random & 0.6001 & 0.3155 & 0.3347 & 0.3441 \\
 &  & SBSampling & 0.7255 & 0.3114 & 0.3421 & 0.3383 \\
 &  & TopStereo & 0.7335 & 0.3243 & 0.3560 & 0.3578 \\
\bottomrule
\end{tabular}
\label{tab:results_overall}
\end{table}
Table~\ref{tab:results_overall} shows the user's gender obfuscation capabilities of \methodours in terms on \bacc for both datasets. Given that both \methodours and the baseline \perblur are independent of the recommendation algorithm, the same values of \bacc are valid for the analysis of the performance of the different recommendation algorithms. We also report the results on the dataset without obfuscations, which we refer to as \textit{original}.
The \bacc values reported correspond to the best values of the average test results over 5-folds for each obfuscation parameter configuration, with the corresponding \ndcg values for each recommendation algorithm, in which at most $10\%$ of the user profiles were obfuscated~(\obfratio$=0.1$).

We observe that \methodours in its variant with \stratremoval and \sbosampling, consistently yields the best results in terms of \bacc for both datasets, proving \methodours's effectiveness in preventing the attacker's ability to infer user's protected attributes, at the cost of a slight decrease in \ndcg
. With \stratremoval and \sbosampling, \methodours delivers $\sim7\%$ and $\sim9\%$ of improvement in \bacc with respect to the original \dataLFM and the original \dataMlm dataset, respectively, at the cost of $\sim2\%$ decrease in \ndcg across all RSs. Furthermore, when compared with \perblur, $\sim8\%$ and $\sim6\%$ in improvement in \bacc is achieved on \dataLFM and \dataMlm, respectively, which translates into  a decrease of $\sim4\%$ in \ndcg on  \dataLFM, and a $\sim1\%$ decrease in \ndcg on  \dataMlm.

We observe that when imputing items, \methodours can have a negative impact on \bacc for most obfuscation configurations; this may be due to the size of the sampling pool. In this regard, \perblur shows more robustness, which might be attributed to filtering items using the user-based KNN recommendation algorithm. This emphasizes the substantial influence of the selection of obfuscation candidates for imputation of user preferences. 


From Table~\ref{tab:results_overall}, we can also speculate that on the original \dataLFM it is already hard to infer the users' gender attribute from their preferences, given the low \bacc values reported. In comparison, \dataMlm is more exposed to adversarial attacks inferring users' gender (higher \bacc on original dataset), and also more sensitive to the obfuscation methods applied, given the fluctuation in the values of \bacc when different obfuscation strategies are used
.  

\begin{figure}[t]
    \centering
    \begin{subfigure}{0.8\textwidth}
    \centering
        \includegraphics[width=\linewidth]{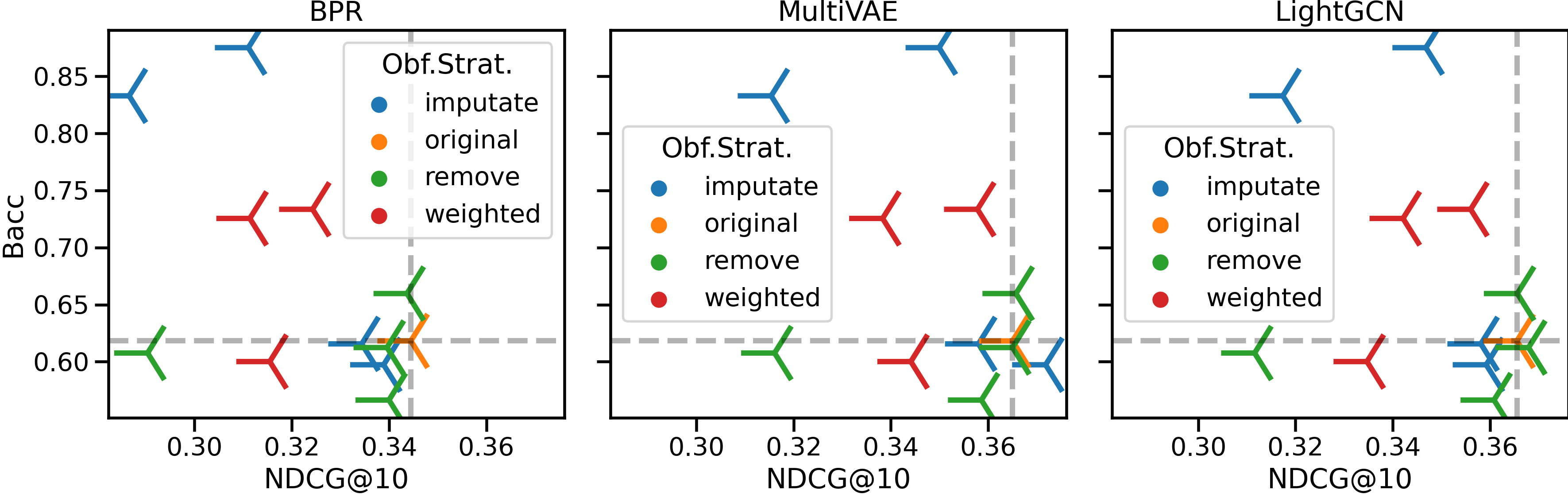}
        \caption{\dataMlm}
        \label{fig:obf_method_a}
    \end{subfigure}
    \begin{subfigure}{0.8\textwidth}
        \centering
         \includegraphics[width=\linewidth]{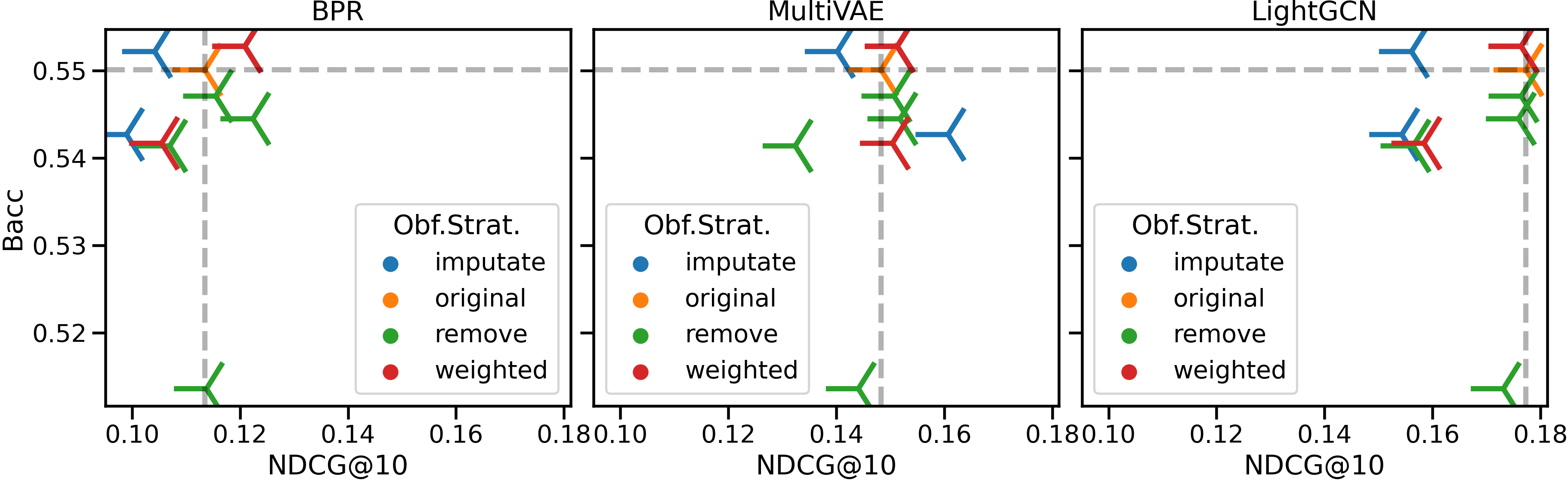}
         \caption{\dataLFM}
         \label{fig:obf_method_b}
    \end{subfigure}
    \caption{Performance of the RSs and attacker (NDCG$@10$ and \bacc) with different obfuscation strategies on (a) \dataMlm and (b) \dataLFM. The dotted lines indicate the performances on the datasets without any obfuscation in place.}
    \label{fig:obf_method_influenc}
\end{figure}
\begin{figure}[t]
    \centering
    \begin{subfigure}{0.8\textwidth}
    \centering
        \includegraphics[width=\linewidth]{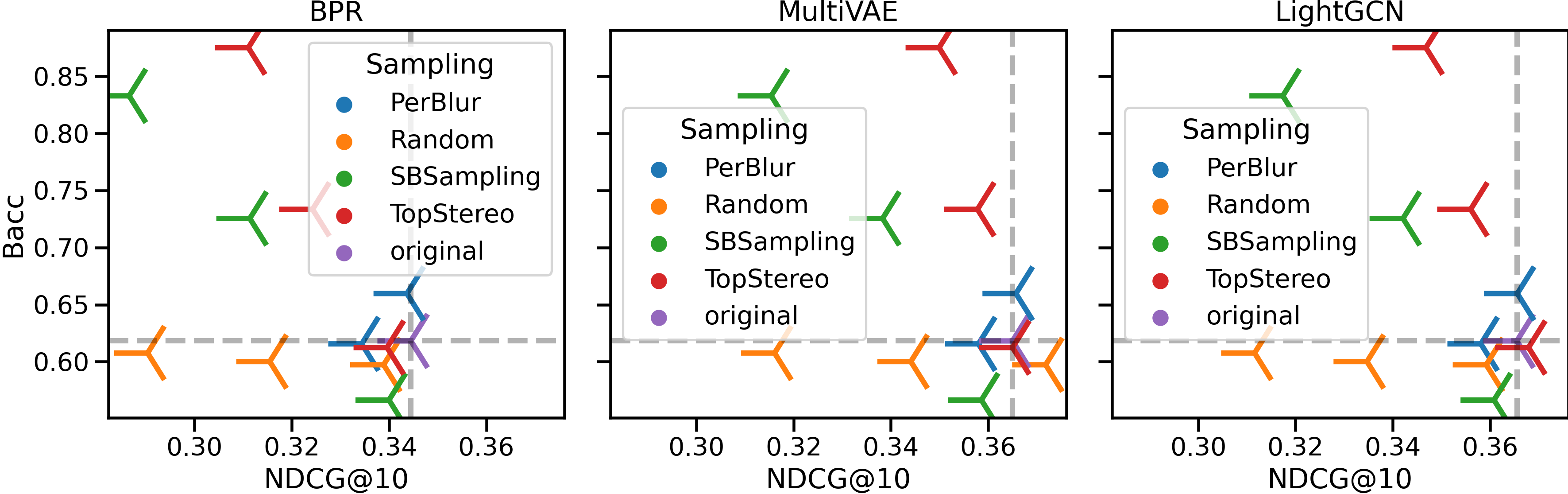}
        \caption{\dataMlm}
        \label{fig:sampl_method_a}
    \end{subfigure}
    \begin{subfigure}{0.8\textwidth}
    \centering
         \includegraphics[width=\linewidth]{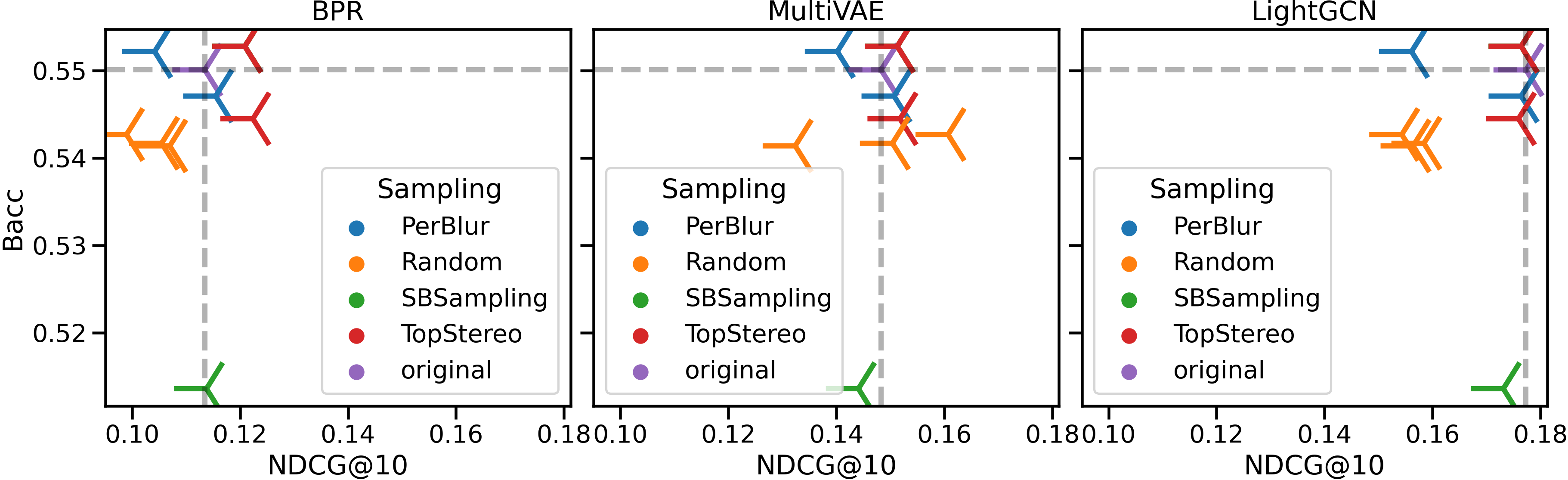}
         \caption{\dataLFM}
         \label{fig:sampl_method_b}
    \end{subfigure}
    \caption{Performance of the RSs and attacker (NDCG$@10$ and \bacc) with different sampling methods on (a) \dataMlm and (b) \dataLFM. The dotted lines indicate the performances on the datasets without any obfuscation in place.
    }
    \label{fig:sampling_method_influenc}
\end{figure}

Figure~\ref{fig:obf_method_influenc} and Figure~\ref{fig:sampling_method_influenc} show the results of the obfuscation strategy and sampling method obfuscation parameters from Table~\ref{tab:results_overall} in terms of two-dimensional plots with \ndcg on the $x$-axis and \bacc on the $y$-axis, and for each recommendation algorithm. In each subplot, the points closer to the bottom-right corner provide better accuracy--privacy trade-off (higher \ndcg and lower \bacc). 

In Figure~\ref{fig:obf_method_influenc}, we see that for both datasets, \stratremoval is usually below the original dataset \bacc values~(below the dotted line), indicating the effectiveness of \stratremoval in preventing adversarial attacks on protected attributes. Other points clearly show improvements in \ndcg, although with a lesser impact on \bacc compared to \stratremoval. The effect of \stratremoval is larger on \dataLFM. Furthermore, for the \stratweight strategy, we observe that the performance of \methodours mostly falls in the central regions of the plots
. Since varying $\omega\in[0, 1]$ allows adjusting the level of  \stratimputate and \stratremoval, we speculate that the parameters of \stratweight could be optimized to target better privacy-oriented results.
Figure~\ref{fig:sampling_method_influenc} compares the performance of \methodours with different sampling methods. We observe that on \dataMlm, \sbosampling and \topstereosampling have decreasing behavior in terms of \bacc while increasing in \ndcg values. On the other hand, \perblur has an ascending tendency. On the \dataLFM dataset, the results are more diverse and only partially resemble the trends observed on \dataMlm. More importantly, the behavior of \sbosampling is similar across different recommendation algorithms.

\section{Conclusion and Future Work}\label{sec:conclusions}
In this work, we introduced \methodours, a novel probabilistic user preference obfuscation method that selects the items to obfuscate based on stereotypicality measures for users and items. 
Our experiments show that \methodours can reach a better accuracy--privacy trade-off than the 
baselines used for comparison on two recommendation domains (music and movies) by removing highly stereotypical items from the users' profiles. In addition, we show that the different configurations of \methodours (obfuscation and sampling strategy)  have similar behavior across different recommendation algorithms.   

In this work, we limited the analysis to gender as the protected attribute and oversimplified its definition, reducing it to a binary attribute. Therefore, we plan to extend the current work by including an analysis of the effect of \methodours with user attributes beyond binary categories, such as age groups or ethnicities. Additionally, our experiments hinted that \stratimputate has the potential to achieve a better accuracy--privacy trade-off, a hypothesis that we leave for future work. 
Finally, further analyses can target the mitigation of other user privacy objectives, such as membership inference. 


\begin{credits}
\subsubsection{\ackname} 
This research was funded in whole or in part by the FFG COMET center program, by the Austrian Science Fund (FWF): P36413, P33526, and DFH-23, and by the State of Upper Austria and the Federal Ministry of Education, Science, and Research, through grants LIT-2021-YOU-215 and LIT-2020-9-SEE-113.
\subsubsection{\discintname}
The authors have no competing interests to declare that are relevant to the content of this article.
\end{credits}


\begin{thebibliography}{10}
\providecommand{\url}[1]{\texttt{#1}}
\providecommand{\urlprefix}{URL }
\providecommand{\doi}[1]{https://doi.org/#1}

\bibitem{anelli2021federank}
Anelli, V.W., Deldjoo, Y., Di~Noia, T., Ferrara, A., Narducci, F.: Federank: User controlled feedback with federated recommender systems. In: Advances in Information Retrieval: 43rd European Conference on IR Research (ECIR 2021). pp. 32--47. Springer (2021)

\bibitem{anelli2022adversarial_handbook}
Anelli, V.W., Deldjoo, Y., Noia, T.D., Merra, F.A.: Adversarial Recommender Systems: Attack, Defense, and Advances, pp. 335--380. Springer US, New York, NY (2022)

\bibitem{deldjoo2022fairness}
Deldjoo, Y., Jannach, D., Bellogin, A., Difonzo, A., Zanzonelli, D.: Fairness in recommender systems: research landscape and future directions. User Modeling and User-Adapted Interaction  \textbf{34}(1) (2024)

\bibitem{deldjoo2021adversarial_fairness}
Deldjoo, Y., Noia, T.D., Merra, F.A.: A survey on adversarial recommender systems: From attack/defense strategies to generative adversarial networks. ACM Comput. Surv.  \textbf{54}(2) (mar 2021). \doi{10.1145/3439729}, \url{https://doi.org/10.1145/3439729}

\bibitem{dwork2008differential}
Dwork, C.: Differential privacy: A survey of results. In: International Conference on Theory and Applications of Models of Computation. pp. 1--19. Springer (2008)

\bibitem{dwork2012fairness}
Dwork, C., Hardt, M., Pitassi, T., Reingold, O., Zemel, R.: Fairness through awareness. In: Proceedings of the 3rd innovations in theoretical computer science conference (ITCS). pp. 214--226 (2012)

\bibitem{ekstrand2022handbook_fairness}
Ekstrand, M.D., Das, A., Burke, R., Diaz, F.: Fairness in Recommender Systems, pp. 603--646. Springer US, New York, NY (2022)

\bibitem{ekstrand2018privacy}
Ekstrand, M.D., Joshaghani, R., Mehrpouyan, H.: Privacy for all: Ensuring fair and equitable privacy protections. In: Conference on fairness, accountability and transparency. pp. 35--47. PMLR (2018)

\bibitem{ganhoer2022advmultvae}
Ganh\"{o}r, C., Penz, D., Rekabsaz, N., Lesota, O., Schedl, M.: Unlearning protected user attributes in recommendations with adversarial training. In: Proceedings of the 45th International ACM SIGIR Conference. p. 2142–2147. SIGIR '22, ACM, New York, NY, USA (2022). \doi{10.1145/3477495.3531820}, \url{https://doi.org/10.1145/3477495.3531820}

\bibitem{DBLP:journals/tiis/HarperK16}
Harper, F.M., Konstan, J.A.: The movielens datasets: History and context. {ACM} Trans. Interact. Intell. Syst.  \textbf{5}(4),  19:1--19:19 (2016). \doi{10.1145/2827872}, \url{https://doi.org/10.1145/2827872}

\bibitem{hashemi2022data}
Hashemi, H., Xiong, W., Ke, L., Maeng, K., Annavaram, M., Suh, G.E., Lee, H.H.S.: Data leakage via access patterns of sparse features in deep learning-based recommendation systems. Workshop on Trustworthy and Socially Responsible Machine Learning (TSRML), in conjunction with the 36th Conference on Neural Information Processing Systems (NeurIPS)  (2022)

\bibitem{DBLP:conf/sigir/LighGCN}
He, X., Deng, K., Wang, X., Li, Y., Zhang, Y., Wang, M.: Lightgcn: Simplifying and powering graph convolution network for recommendation. In: Huang, J.X., Chang, Y., Cheng, X., Kamps, J., Murdock, V., Wen, J., Liu, Y. (eds.) Proceedings of the 43rd International {ACM} {SIGIR} conference on research and development in Information Retrieval, {SIGIR} 2020, Virtual Event, China, July 25-30, 2020. pp. 639--648. {ACM} (2020). 

\bibitem{jin2023fairness}
Jin, D., Wang, L., Zhang, H., Zheng, Y., Ding, W., Xia, F., Pan, S.: A survey on fairness-aware recommender systems. Information Fusion  \textbf{100},  101906 (2023). 

\bibitem{kim2016efficient}
Kim, S., Kim, J., Koo, D., Kim, Y., Yoon, H., Shin, J.: Efficient privacy-preserving matrix factorization via fully homomorphic encryption. In: Proceedings of the 11th ACM on Asia conference on computer and communications security (ASIACCS). pp. 617--628 (2016)

\bibitem{DBLP:journals/mta/KrismayerSKR19}
Krismayer, T., Schedl, M., Knees, P., Rabiser, R.: Predicting user demographics from music listening information. Multim. Tools Appl.  \textbf{78}(3),  2897--2920 (2019). \doi{10.1007/S11042-018-5980-Y}, \url{https://doi.org/10.1007/s11042-018-5980-y}

\bibitem{lacic2020using}
Lacic, E., Reiter-Haas, M., Kowald, D., Reddy~Dareddy, M., Cho, J., Lex, E.: Using autoencoders for session-based job recommendations. User Modeling and User-Adapted Interaction  \textbf{30},  617--658 (2020)

\bibitem{lex2020modeling}
Lex, E., Kowald, D., Schedl, M.: Modeling popularity and temporal drift of music genre preferences. Transactions of the International Society for Music Information Retrieval  \textbf{3}(1),  17--31 (2020)

\bibitem{Li2021counterFactMultfairness}
Li, Y., Chen, H., Xu, S., Ge, Y., Zhang, Y.: Towards personalized fairness based on causal notion. In: Proceedings of the 44th International ACM SIGIR Conference on Research and Development in Information Retrieval. p. 1054–1063. SIGIR '21, ACM, New York, NY, USA (2021). 

\bibitem{Li2023IndistinguishableUsers}
Li, Y., Chen, C., Zheng, X., Zhang, Y., Han, Z., Meng, D., Wang, J.: Making users indistinguishable: Attribute-wise unlearning in recommender systems. In: Proceedings of the 31st ACM International Conference on Multimedia. p. 984–994. MM '23, ACM, New York, NY, USA (2023). 

\bibitem{LiangMultVAE2018}
Liang, D., Krishnan, R.G., Hoffman, M.D., Jebara, T.: Variational autoencoders for collaborative filtering. In: Proceedings of the 2018 World Wide Web Conference. p. 689–698. WWW '18, International World Wide Web Conferences Steering Committee, Republic and Canton of Geneva, CHE (2018). 

\bibitem{DBLP:conf/trustcom/LinLZWHL22}
Lin, C., Liu, B., Zhang, X., Wang, Z., Hu, C., Luo, L.: Privacy-preserving recommendation with debiased obfuscaiton. In: {IEEE} International Conference on Trust, Security and Privacy in Computing and Communications, TrustCom 2022, Wuhan, China, December 9-11, 2022. pp. 590--597. {IEEE} (2022). 

\bibitem{lin2020meta}
Lin, Y., Ren, P., Chen, Z., Ren, Z., Yu, D., Ma, J., Rijke, M.d., Cheng, X.: Meta matrix factorization for federated rating predictions. In: Proceedings of the 43rd International ACM SIGIR Conference on Research and Development in Information Retrieval (SIGIR). pp. 981--990. Springer (2020)

\bibitem{melchiorre2021gender_fairness}
Melchiorre, A.B., Rekabsaz, N., Parada{-}Cabaleiro, E., Brandl, S., Lesota, O., Schedl, M.: Investigating gender fairness of recommendation algorithms in the music domain. Inf. Process. Manag.  \textbf{58}(5),  102666 (2021). 

\bibitem{muellner2021robustness}
Muellner, P., Kowald, D., Lex, E.: Robustness of meta matrix factorization against strict privacy constraints. In: European Conference on Information Retrieval. pp. 107--119 (2021)

\bibitem{reuseknn}
M\"{u}llner, P., Lex, E., Schedl, M., Kowald, D.: Reuseknn: Neighborhood reuse for differentially-private knn-based recommendations. ACM Trans. Intell. Syst. Technol.  (2023). 

\bibitem{mullner2024impact}
M{\"u}llner, P., Lex, E., Schedl, M., Kowald, D.: The impact of differential privacy on recommendation accuracy and popularity bias. In: European Conference on Information Retrieval. pp. 466--482. Springer (2024)

\bibitem{Mullner2023DP-CF}
Müllner, P., Lex, E., Schedl, M., Kowald, D.: Differential privacy in collaborative filtering recommender systems: a review. Frontiers in Big Data  \textbf{6} (2023). 

\bibitem{DBLP:conf/uai/RendleFGS09}
Rendle, S., Freudenthaler, C., Gantner, Z., Schmidt-Thieme, L.: Bpr: Bayesian personalized ranking from implicit feedback. In: Proc. of UAI. pp. 452--461 (2009)

\bibitem{DBLP:journals/ijmir/Schedl17}
Schedl, M.: Investigating country-specific music preferences and music recommendation algorithms with the lfm-1b dataset. Int. J. Multim. Inf. Retr.  \textbf{6}(1),  71--84 (2017). 

\bibitem{Schedl2022LFM2b}
Schedl, M., Brandl, S., Lesota, O., Parada-Cabaleiro, E., Penz, D., Rekabsaz, N.: Lfm-2b: A dataset of enriched music listening events for recommender systems research and fairness analysis. In: Proceedings of the 2022 Conference on Human Information Interaction and Retrieval. p. 337–341. CHIIR '22, ACM, New York, NY, USA (2022). 

\bibitem{DBLP:journals/ipm/SlokomHL21}
Slokom, M., Hanjalic, A., Larson, M.A.: Towards user-oriented privacy for recommender system data: {A} personalization-based approach to gender obfuscation for user profiles. Inf. Process. Manag.  \textbf{58}(6),  102722 (2021). 

\bibitem{DBLP:conf/recsys/StrucksSL19}
Strucks, C., Slokom, M., Larson, M.A.: Blurm(or)e: Revisiting gender obfuscation in the user-item matrix. In: Burke, R., Abdollahpouri, H., Malthouse, E.C., Thai, K.P., Zhang, Y. (eds.) Proceedings of the Workshop on Recommendation in Multi-stakeholder Environments co-located with the 13th {ACM} Conference on Recommender Systems (RecSys 2019), Copenhagen, Denmark, September 20, 2019. {CEUR} Workshop Proceedings, vol.~2440. CEUR-WS.org (2019).

\bibitem{DBLP:conf/recsys/VassoyLK23}
Vass{\o}y, B., Langseth, H., Kille, B.: Providing previously unseen users fair recommendations using variational autoencoders. In: Zhang, J., Chen, L., Berkovsky, S., Zhang, M., Noia, T.D., Basilico, J., Pizzato, L., Song, Y. (eds.) Proceedings of the 17th {ACM} Conference on Recommender Systems, RecSys 2023, Singapore, Singapore, September 18-22, 2023. pp. 871--876. {ACM} (2023). 

\bibitem{verma2018fairness}
Verma, S., Rubin, J.: Fairness definitions explained. In: Proceedings of the international workshop on software fairness. pp.~1--7 (2018)

\bibitem{Wang23SurveyFairness}
Wang, Y., Ma, W., Zhang, M., Liu, Y., Ma, S.: A survey on the fairness of recommender systems. ACM Trans. Inf. Syst.  \textbf{41}(3) (feb 2023). 

\bibitem{weinsberg2012blurme}
Weinsberg, U., Bhagat, S., Ioannidis, S., Taft, N.: Blurme: Inferring and obfuscating user gender based on ratings. In: Proceedings of the sixth ACM conference on Recommender systems. pp. 195--202 (2012)

\bibitem{wu2021adversarial_fairness}
Wu, C., Wu, F., Wang, X., Huang, Y., , Xie, X.: Fairness-aware news recommendation with decomposed adversarial learning. In: Proc. of AAAI Conference on Artificial Intelligence. p. 4462–4469 (2021)

\bibitem{wu2021representation_fairness}
Wu, L., Chen, L., Shao, P., Hong, R., Wang, X., Wang, M.: Learning fair representations for recommendation: A graph-based perspective. In: Proceedings of the Web Conference 2021. p. 2198–2208. WWW '21, ACM, New York, NY, USA (2021). 

\bibitem{wu2023fairness}
Wu, Y., Cao, J., Xu, G.: Fairness in recommender systems: Evaluation approaches and assurance strategies. ACM Trans. Knowl. Discov. Data  \textbf{18}(1) (aug 2023). 

\bibitem{xin2023user}
Xin, X., Yang, J., Wang, H., Ma, J., Ren, P., Luo, H., Shi, X., Chen, Z., Ren, Z.: On the user behavior leakage from recommender system exposure. ACM Transactions on Information Systems (TOIS)  \textbf{41}(3),  1--25 (2023)

\bibitem{zemel2013learning}
Zemel, R., Wu, Y., Swersky, K., Pitassi, T., Dwork, C.: Learning fair representations. In: International conference on machine learning (ICML). pp. 325--333 (2013)

\bibitem{zhang2021privacy}
Zhang, M., Chen, Y., Lin, J.: A privacy-preserving optimization of neighborhood-based recommendation for medical-aided diagnosis and treatment. IEEE Internet of Things Journal  \textbf{8}(13),  10830--10842 (2021)

\bibitem{zhang2022comprehensive}
Zhang, S., Yin, H.: Comprehensive privacy analysis on federated recommender system against attribute inference attacks. IEEE Transactions on Knowledge and Data Engineering (TKDE)  (2023)

\end{thebibliography}

\appendix
\section{Appendix}\label{sec:appendix1}
\begin{table}[h]
\caption{Experimental results on the two datasets \dataMlm and \dataLFM. The scores in \textbf{bold} indicate the best scores across all models using the obfuscation ratio \obfratio$=0.05$.}
\centering
\setlength{\tabcolsep}{6pt}
\begin{tabular}{lllrrrr}
\toprule
 &  &  &  & \modelbpr & \modellightgcn & \modelmultvae \\
 \cmidrule{5-7}
Dataset& Obf. Strat.  & Sampling &\multicolumn{1}{c}{\bacc$\downarrow$} & \multicolumn{3}{c}{\ndcg$\uparrow$} \\
\midrule
 & original & original & 0.5501 & 0.1135 & 0.1773 & 0.1483 \\
\cdashline{3-7}
\multirow[c]{10}{*}{\dataLFM} & \multirow[c]{3}{*}{imputate} & Random & 0.5464 & 0.1081 & 0.1680 & \textbf{0.1564 }\\
 &  & SBSampling & 0.5528 & 0.1209 & 0.1764 & 0.1513 \\
 &  & TopStereo & 0.5528 & 0.1209 & 0.1764 & 0.1513 \\
\cdashline{3-7}
 & \multirow[c]{3}{*}{remove} & Random & 0.5258 & 0.1132 & 0.1684 & 0.1402 \\
 &  & SBSampling & \textbf{0.5184} & 0.1195 & \textbf{0.1774} & 0.1448 \\
 &  & TopStereo & 0.5445 & \textbf{0.1224} & 0.1759 & 0.1518 \\
 \cdashline{3-7}
 & \multirow[c]{3}{*}{weighted} & Random & 0.5385 & 0.1154 & 0.1710 & 0.1513 \\
 &  & SBSampling & 0.5528 & 0.1209 & 0.1764 & 0.1513 \\
 &  & TopStereo & 0.5528 & 0.1209 & 0.1764 & 0.1513 \\
 \midrule
 & original & original & 0.6182 & \textbf{0.3445} & 0.3655 & 0.3650 \\
\cdashline{3-7}
\multirow[c]{10}{*}{\dataMlm} & \multirow[c]{3}{*}{imputate} & Random & 0.6017 & 0.3429 & 0.3628 & \textbf{0.3702 }\\
 &  & SBSampling & 0.7361 & 0.3127 & 0.3438 & 0.3419 \\
 &  & TopStereo & 0.7335 & 0.3243 & 0.3560 & 0.3578 \\
\cdashline{3-7}
 & \multirow[c]{3}{*}{remove} & Random & 0.6044 & 0.3177 & 0.3376 & 0.3372 \\
 &  & SBSampling &\textbf{ 0.5794} & 0.3443 & 0.3647 & 0.3639 \\
 &  & TopStereo & 0.6124 & 0.3396 & \textbf{0.3679}& 0.3650 \\
 \cdashline{3-7}
 & \multirow[c]{3}{*}{weighted} & Random & 0.6142 & 0.3290 & 0.3543 & 0.3560 \\
 &  & SBSampling & 0.67614 & 0.3259 & 0.3592 & 0.3583 \\
 &  & TopStereo & 0.6692 & 0.3282 & 0.3572 & 0.3637 \\
\bottomrule
\end{tabular}
\end{table}

\begin{figure}[h]
       \centering
    \begin{subfigure}{0.8\textwidth}
    \centering
        \includegraphics[width=\linewidth]{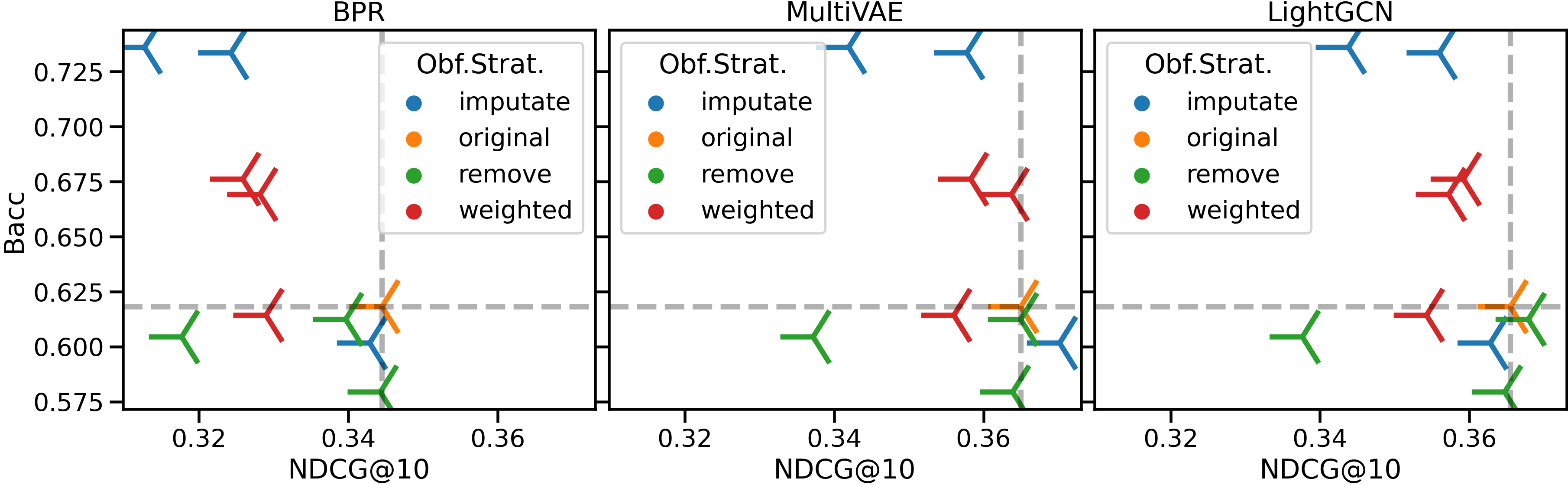}
        \caption{\dataMlm}
        \label{fig:sup_obf_method_a}
    \end{subfigure}
    \begin{subfigure}{0.8\textwidth}
        \centering
         \includegraphics[width=\linewidth]{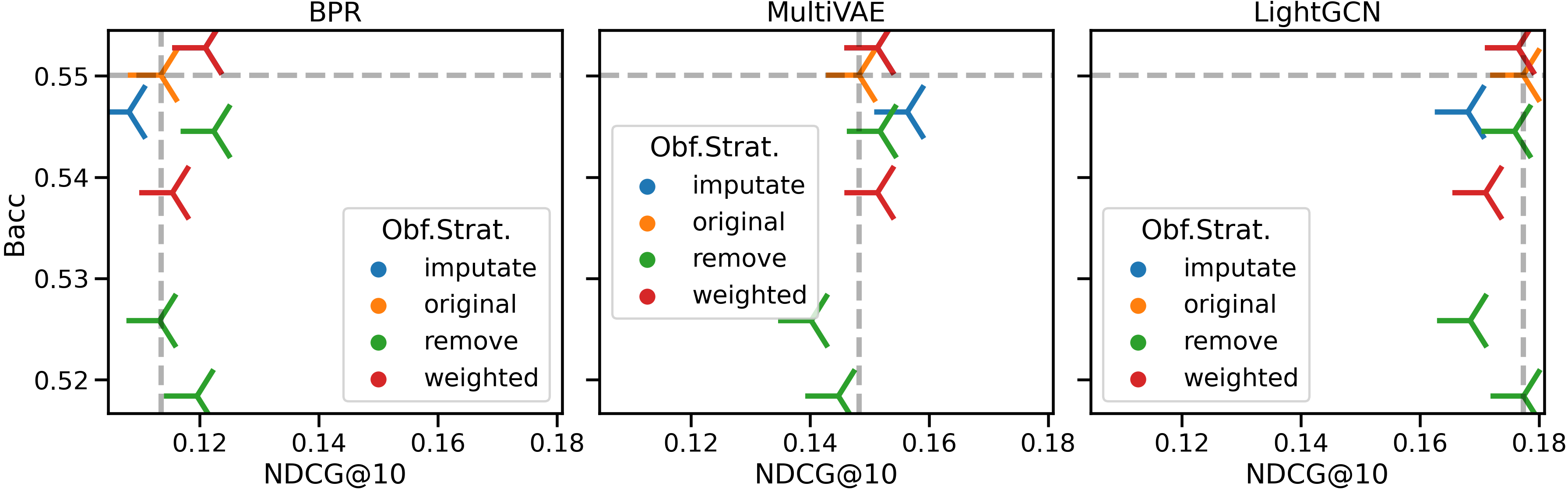}
         \caption{\dataLFM}
         \label{fig:sup_obf_method_b}
    \end{subfigure}
    \caption{Performance of the RSs and attacker (NDCG$@10$ and \bacc) with different obfuscation strategies on (a) \dataMlm and (b) \dataLFM using the obfuscation ratio \obfratio$=0.05$. The dotted lines indicate the performances on the datasets without any obfuscation in place.}

\end{figure}
\begin{figure}[h]
    \centering
    \begin{subfigure}{0.8\textwidth}
    \centering
        \includegraphics[width=\linewidth]{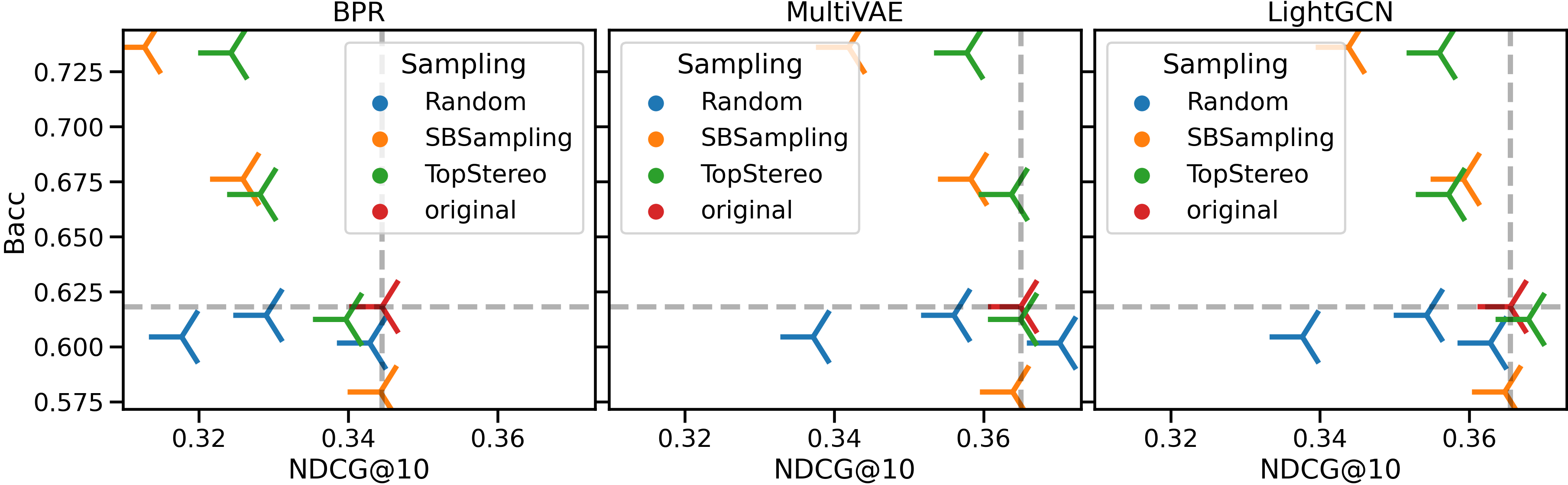}
        \caption{\dataMlm}
        \label{fig:sup_sampl_method_a}
    \end{subfigure}
    \begin{subfigure}{0.8\textwidth}
    \centering
         \includegraphics[width=\linewidth]{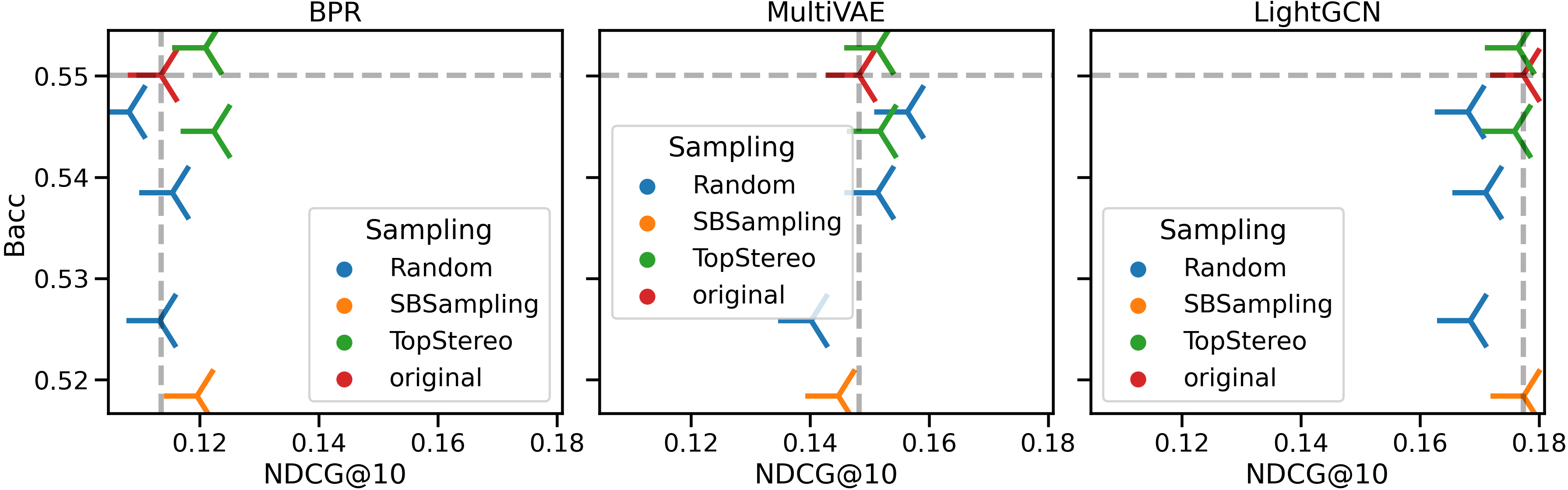}
         \caption{\dataLFM}
         \label{fig:sup_sampl_method_b}
    \end{subfigure}
\caption{Performance of the RSs and attacker (NDCG$@10$ and \bacc) with different sampling methods on (a) \dataMlm and (b) \dataLFM using the obfuscation ratio \obfratio$=0.05$. The dotted lines indicate the performances on the datasets without any obfuscation in place.}
\end{figure}

\end{document}
\endinput

\section{Research Methods}

\subsection{Part One}

Lorem ipsum dolor sit amet, consectetur adipiscing elit. Morbi
malesuada, quam in pulvinar varius, metus nunc fermentum urna, id
sollicitudin purus odio sit amet enim. Aliquam ullamcorper eu ipsum
vel mollis. Curabitur quis dictum nisl. Phasellus vel semper risus, et
lacinia dolor. Integer ultricies commodo sem nec semper.

\subsection{Part Two}

Etiam commodo feugiat nisl pulvinar pellentesque. Etiam auctor sodales
ligula, non varius nibh pulvinar semper. Suspendisse nec lectus non
ipsum convallis congue hendrerit vitae sapien. Donec at laoreet
eros. Vivamus non purus placerat, scelerisque diam eu, cursus
ante. Etiam aliquam tortor auctor efficitur mattis.

\section{Online Resources}

Nam id fermentum dui. Suspendisse sagittis tortor a nulla mollis, in
pulvinar ex pretium. Sed interdum orci quis metus euismod, et sagittis
enim maximus. Vestibulum gravida massa ut felis suscipit
congue. Quisque mattis elit a risus ultrices commodo venenatis eget
dui. Etiam sagittis eleifend elementum.
Quiero que me trates suavemente
Nam interdum magna at lectus dignissim, ac dignissim lorem
rhoncus. Maecenas eu arcu ac neque placerat aliquam. Nunc pulvinar
massa et mattis lacinia.

\end{document}